\documentclass[
prd,
superscriptaddress,
10 pt,
preprintnumbers,
notitlepage,
secnumarabic,
nobibnotes,
nofootinbib,
showpacs
]{revtex4-1}

\usepackage[T1]{fontenc}
\usepackage{bm}
\usepackage[pdftex]{color,graphicx}
\usepackage[makeroom]{cancel}
\usepackage{amssymb}
\usepackage{mathtools}
\usepackage{tabularx}
\usepackage{dsfont}
\usepackage{nicefrac}
\usepackage{hyperref}
\usepackage[artemisia]{textgreek}
\usepackage{tikz}
\usetikzlibrary{shapes}
\usetikzlibrary{arrows}
\tikzset{l/.style={draw=black, line width=1pt}}
\tikzset{v/.style={auto, outer sep=-2}}
\usepackage{perpage}
\MakePerPage{footnote}
\newcommand{\kbar}{\mathchar'26\mkern-9mu k}
\newcommand{\bra}[1]{\ensuremath{\left\langle#1\right|}}
\newcommand{\ket}[1]{\ensuremath{\left|#1\right\rangle}}
\newcommand{\bracket}[2]{\ensuremath{\left\langle #1 \middle| #2 \right\rangle}}

\newcommand{\vket}[1]{\ensuremath{\Biggl|\vcenter{\hbox{#1}}\Biggr\rangle}}

\DeclareMathAlphabet\mathbfcal{OMS}{cmsy}{b}{n}

\begin{document}
\title{2+1 homogeneous Loop Quantum Gravity with a scalar field clock}

\author{Jakub Bilski}
\email{jakubbilski14@fudan.edu.cn}
\affiliation{Center for Field Theory and Particle Physics \& Department of Physics, Fudan University, 200433 Shanghai, China}

\author{Antonino Marcian\`o} 
\email{marciano@fudan.edu.cn}
\affiliation{Center for Field Theory and Particle Physics \& Department of Physics, Fudan University, 200433 Shanghai, China}

	%%%%%%%%%	%%%%%%%%%	%%%%%%%%%	%%%%%%%%%

	%%%%%%%%%	%%%%%%%%%	%%%%%%%%%	%%%%%%%%%
\begin{abstract}
\noindent
We focus on three-dimensional QRLG with the purpose of shedding light on the link between reduced LQG and LQC in four space-time dimensions. Considering homogeneous three-dimensional LQG, the theory simplifies to QRLG. We then implement Thiemann's Quantum Spin Dynamics for Euclidean three-dimensional space-time in presence of a real scalar matter field. We deploy a polymer quantization of the scalar field while using methods of quantum reduced loop gravity. We compute the scalar Hamiltonian operator on the states of the kinematical Hilbert space of the theory, and exhibit its matrix elements that are derived using a new simplified method. The coupling to matter, which plays the role of a carrier of dynamics, opens the pathway to the study of phenomenological implications. We finally comment on the relations between three-dimensional QRLG and LQC, as well as on the appearance of the correspondence principle for the scalar field. 
\end{abstract}
	%%%%%%%%%	%%%%%%%%%	%%%%%%%%%	%%%%%%%%%
\maketitle

	%%%%%%%%%	%%%%%%%%%	%%%%%%%%%	%%%%%%%%%
	
	%%%%%%%%%	%%%%%%%%%	%%%%%%%%%	%%%%%%%%%
\section{Introduction}\label{I}
\noindent
Loop Quantum Gravity (LQG) \cite{Ashtekar:2004eh,Rovelli:2004tv,Thiemann:2007zz} is an attempt of
background independent quantization of Einsteinian gravity. The theory cast 
in the first order formalism may be reshuffled --- in the original proposal by Ashtekar --- in terms of variables with a manifest internal SU$(2)$ gauge symmetry. Canonical SU$(2)$ variables are then replaced by smeared quantities, which can be quantized deploying Wilsonian loop methods firstly developed for lattice gauge theories' quantization \cite{Rovelli:2004tv}.

The strategy adopted by LQG is technically very neat and conceptually very clear. Nonetheless, after almost thirty years it remains challenging to accomplish the quantization of the full theory in $3+1$ dimensions --- alternative ways have been suggested \cite{Mercati:2014ama}, while it was also pointed out that this difficultly might eventually hide a conceptual flaw in the general approach of canonical quantization \cite{Murchadha:2012zz}. Conversely, the investigation of lower dimensional cases shows, within a fundamentally exemplified version of the problem, surprising results. Much attention on those latter was paid with the purpose of extracting a useful intuition how to extend techniques successful in the $3+1$ dimensional case. Indeed, it is widely renown that quantum gravity in $2+1$ dimensions is an exact soluble problem, with extensive studies been devoted to this topic from the last decades --- see \textit{e.g.} the classical Ref.~\cite{Witten:1988hc}. Studies on some particular models of LQG suggest that this theory should be also exactly solvable in three dimensions \cite{Alesci:2013xd}. The role of the cosmological constant in modifying the internal symmetries of the theory at the quantum level was then deepened for the first time in \cite{Witten:1988hf}. From the perspective of LQG, the lower dimensional case was studied by Thiemann in Ref.~\cite{Thiemann:1997ru} --- we refer also the reader to a series of studies by Noui and Perez \cite{NoPe,Noui:2004iz}. The loop quantization of the $2+1$ dimensional Einsteinian theory of gravity with cosmological constant is still under scrutiny, in order to check whether the state sum encoded in the Turaev-Viro \cite{TV} model can be recovered. Finally, the three-dimensional model of Loop Quantum Cosmology (LQC) was analyzed first in the isotropic case \cite{Zhang:2014xqa}, then later extended to the anisotropic one \cite{Ding:2016spw}. Authors demonstrated how the classical cosmological singularity is replaced by a quantum bouncing evolution, reproducing the main results of full LQC.
\\
 
Coming back to the $3+1$ dimensional theory, symmetry reduced models were proposed as a simplified framework to be studied, in order to test quantizations' techniques of LQG and capture the cosmological sector of the theory --- black hole metrics were also captured and dealt with in this symmetry-reduction scheme. Historically, the first attempt of symmetry reduction was LQC, for which we refer to \cite{Bojowald:2011zzb,Ashtekar:2011ni,Banerjee:2011qu}. Several other possibilities then followed, with the clear purpose of linking the full theory of LQG to its cosmological sector. For instance, the embedding of the quantum configuration spaces of LQC into the full theory was deepened in Ref.~\cite{Brunnemann:2007du, Engle:2013qq, Hanusch:2013jza, Fleischhack:2015nda}, while the use of spinfoam techniques was investigated in \cite{Rovelli:2009tp}. Coherent state techniques were then proposed from a Group Field Theory perspective in \cite{Gielen:2013naa,Oriti:2016qtz}. Gauge fixing procedures were only recently proposed. The first model was introduced in Refs.~\cite{Alesci:2012md,Alesci:2013xd} and dubbed Quantum Reduced Loop Gravity (QRLG) --- see also \cite{Alesci:2016gub} for a review of QRLG and \cite{Alesci:2014uha,Bilski:2016pib} for developments. Then in \cite{Bodendorfer:2014vea} it was proposed an unfixing procedure very close to the implementation of Dirac brackets, and the quantization of a reduced phase space classically gauge fixed.

Specifically, QRLG implements weakly gauge-fixing to the states of the kinematical Hilbert space of LQG. The classical equivalent of imposing weakly these quantum constraints amounts to gauge-fix the spatial part of the metric and the dreibein to a diagonal form. The formalism hitherto developed allowed to consistently address the quantization of Bianchi I models, and to recover within the semiclassical limit the effective Hamiltonian of LQC \cite{Alesci:2013xd,Alesci:2014uha} in the $\mu_0$ regularization scheme. An effective improved dynamics \cite{Ashtekar:2006wn} can be obtained too by averaging over the ensemble of the classically equivalent states \cite{Alesci:2014rra,Alesci:2015nja,Alesci:2016rmn}.

The main advantage of QRLG stands not only on a novel derivation of results already obtained in LQC --- see \textit{e.g.} the realization of the bouncing scenario \cite{Alesci:2013xd,Alesci:2014rra} --- but rather in the development of a quantized matter sector (on a quantized background independent scheme) reproducing the Standard Model of particle physics. This represents a clear advancement for the full theory. Matter fields' quantization is achieved via the tools of the loop quantization \cite{Thiemann:1997rq,Thiemann:1997rt}. This is consistent with the very fundamental ideas of QRLG. The latter offers a framework in which we can test implications of loop quantization for matter fields, and hence develop a phenomenology for this sector. The first analysis focusing on a scalar matter field was developed in \cite{Bilski:2015dra}, while the way vector fields can be encoded in QRLG was studied\footnote{Vector fields may have an important role in cosmology \cite{Ford:1989me, Parker:1993ya, Golovnev:2008cf, Maleknejad:2011sq, Alexander:2011hz, Adshead:2012kp, Alexander:2014uza}, and be relevant for the phenomenology of quantum gravity and for development of unified formalisms for all forces \cite{Alexander:2011jf, Alexander:2012ge}.} in \cite{Bilski:2016pib}. The analysis of the quantization of spinor matter fields is currently in progress \cite{fermions}. \\

In this work we develop the quantum dynamics of Einsteinian gravity in $2+1$ dimensions, adding a scalar degree of freedom to the topological lower dimensional set-up. The theory is then analyzed through the lens of QRLG and adopting a polymer quantization of the scalar field. More in details, in Sec.~\eqref{II} we introduce methods used in this paper and provide a short review of LQG and QRLG. In Sec.~\eqref{III} we present the quantum calculations, which are the main content of this study. Specifically, after having specified the use of Wick rotation, in Sec.~\eqref{II.1} we restore SU$(2)$ symmetry also for the three-dimensional theory of gravity. We then obtain a total Hamiltonian being a set of constraints that can be solved providing the kinematical space quantized {\`a la} LQG. This theory has been introduced in the section \eqref{II.2}, where we also summarized QRLG. The latter appears as a natural exemplification to the two-dimensional space, with no dynamical degrees of freedom. Dynamics has been added through coupling matter scalar field \eqref{II.3}, which at the quantum level has been described in a simple polymer model. The key derivation begins with lattice regularization of scalar constraint \eqref{III.1} that has been quantized in the next section \eqref{III.2}, giving an analytical expression for the Hamiltonian operator. It is worth mentioning that it clearly coincides with the Hamiltonian operator of $2+1$ dimensional LQC. Finally we have shown how our results reproduce classical expressions at the leading order \eqref{III.3}. We conclude in Sec.~\eqref{IV} discussing the results we obtained. Some new constructions and methods typical to QRLG has been attached in the Appendix.\\

Through the paper we focus on the Euclidean three-dimensional metric signature $(+,+,+)$. We denote the gravitational coupling constant as $\kappa=16\pi G$, the reduced gravitational coupling constant as $\kbar=\frac{1}{2}\gamma\hbar\kappa=8\pi\gamma l_P^2$ --- $\gamma$ and $l_P$ are the Immirzi parameter and the Planck length respectively --- and denote with $\lambda$ the scalar field coupling constant of dimension $\hbar/l_P^2$. We work in natural units and set the values of the fundamental constant to one, namely $\hbar=c=1$. We use the symbol $'$ to denote variables in the spacetime with Lorentzian signature $(+,+,-)$. The metric tensor is defined as $g_{\mu\nu}=e^i_{\mu}e^i_{\nu}$, where $e^i_{\mu}$ stand for the inverse dreibein fields. Greek letters from the middle of the alphabet: $\mu,\nu,...=x,y,t$ denote Euclidean coordinate indices. Lowercase Latin letters $a,b,...=1,2$ label coordinate on each Cauchy hypersurface constructed by the three-dimensional equivalent to the ADM decomposition \cite{Arnowitt:1962hi}, while $i,j,..=1,2,3$ are $\mathfrak{su}(2)$ internal indices. Indices written in the bracket $``(\,)''$ are not summed, while for for repeated pairs of indices the Einstein convention is applied.

	%%%%%%%%%	%%%%%%%%%	%%%%%%%%%	%%%%%%%%%

	%%%%%%%%%	%%%%%%%%%	%%%%%%%%%	%%%%%%%%%
\section{Loop framework for quantum gravity and scalar field}\label{II}
\noindent
In this section we specify the methodology we deployed during our analysis, an at the same time provide a short review of LQG and QRLG.

\subsection{SU(2) symmetry in three-dimensional gravity}\label{II.1}
\noindent
In analogy to the case of four-dimensional gravity, we describe a three-dimensional model by introducing the action \cite{Witten:1988hc}
\begin{equation}\label{3D_action}
S^{(gr)}:=\frac{1}{\kappa\lambda}\int_{M}\!\!\!\!d^3x\sqrt{|g'|}{R'}^{(3)}
=\frac{1}{4\kappa\lambda}\int_{M}\!\!\!\!d^3x\ \epsilon^{\mu\nu\rho}\epsilon_{ijk}{e'}^{i}_{\mu}
\big(\partial_{\nu}{\Gamma'}^{jk}_{\ \ \rho}-\partial_{\rho}{\Gamma'}^{jk}_{\ \ \nu}
+{\Gamma'}^{jl}_{\ \ \nu}{\Gamma'}^{kl}_{\ \ \rho}-{\Gamma'}^{jl}_{\ \ \rho}{\Gamma'}^{kl}_{\ \ \nu}\big)\,,
\end{equation}
where ${R'}^{(3)}$ is the scalar curvature and $g'$ is the determinant of three-dimensional metric tensor $g'_{\mu\nu}$. It is worth noting that all the expressions are cast in units of Newton's constant $G=l_P^2/\hbar$ --- for simplicity we rescaled the total action by $1/\lambda$, which stands for the inverse of the scalar field coupling constant.

The ADM decomposition \cite{Arnowitt:1962hi} performed on the three-dimensional equivalent of the Minkowski spacetime --- with Lorentzian signature $(-,+,+)$ --- results in the line element
\begin{equation}\label{ADM}
ds^2=q'_{ab}dx^adx^b+2N'_adx^adt-\big({N'}^2-N'_aN'^a\big)dt^2\,,
\end{equation}
where $N'$ and $N'^a$ are the physical lapse function and the shift vector, respectively. Note that the explicit forms of the dreibein and its inverse are 
\begin{subequations}
\\ \noindent\begin{minipage}{0.33 \linewidth}
\begin{align}\label{matrix'1}
{e'}_i^{\mu}=\left( \begin{array}{cc} & 0 \\ {e'}_i^a & 0 \\ & \frac{1}{N'} \end{array} \right),
\end{align}
\end{minipage}
\begin{minipage}{0.33 \linewidth}
\begin{align}\label{matrix'2}
{e'}^i_{\mu}=\left( \begin{array}{ccc} & \!{e'}^i_a &  \\ {e'}^1_aN'^a & {e'}^2_aN'^a & N' \end{array} \right),
\end{align}
\end{minipage}
\begin{minipage}{0.33 \linewidth}
\begin{align}\label{matrix'3}
g'_{\mu\nu}={e'}^i_{\mu}{e'}^i_{\nu},
\end{align}
\end{minipage}\\ \\
\end{subequations}
where the submatrices are ${e'}_i^a=\left({e'}_1^a,{e'}_2^a,-\frac{N'^a}{N'}\right)^{\!T}$, ${e'}^i_a=\left({e'}^1_a,{e'}^2_a,0\right)$, while the 2-metric tensor reads $q'_{ab}={e'}_a^i{e'}_b^i$.

In order to recover the internal SU$(2)$ symmetry instead of SU$(1,1)$ on a spatial subspace of spacetime, one has to perform the internal Wick rotation:
${e'}^3_{\mu}\rightarrow e^3_{\mu}\!=\!-{e'}^3_{\mu}$
and
${e'}_3^{\mu}\rightarrow e_3^{\mu}\!=\!-{e'}_3^{\mu}$.
The physical consequence of this transformation is a reverse of the sign of the lapse function, precisely:
$N'\!=\!-N,\ {N'}^a\!=\!N^a,\ {e'}^i_a\!=\!e^i_a,\ {e'}_i^a\!=\!e_i^a$, while the three-dimensional metric on the spacial Cauchy hypersurface $\Sigma_t$, $q_{ab}=q'_{ab}$ remains unchanged.\\

We can now construct variables in analogy to what has been done by Ashtekar in four-dimensional gravity \cite{Ashtekar:1986yd}. We can define the quantity $\pi^{\ \ \mu}_{ij}:=\kappa\lambda\frac{\partial L^{(gr)}}{\partial(\partial_t\Gamma^{ij}_{\ \ \mu})}$, representing momentum canonically conjugated to the spin connection. It is worth noting that $\pi^{\ \ t}_{ij}=0$, hence $\Gamma^{ij}_{\ \ t}$ is a Lagrange multiplier and the momentum reads $\pi^{\ \ a}_{ij}=\frac{1}{2}\epsilon_{ijk}\epsilon^{ab}e^k_b$, where $\epsilon^{ab}:=\epsilon^{abt}$. Then, the natural choice for a canonical gauge field, similar to the Ashtekar one is
\begin{equation}\label{Ashtekarconnection}
A^i_a:=\frac{1}{2}\epsilon^{ijk}\Gamma^{jk}_{\ \ a}.
\end{equation}
Proceeding in a similar way as LQG, the densitized zweibein canonically conjugated to $A^i_a$ becomes
\begin{equation}\label{DensitizedZweibein}
E_i^a:=\epsilon^{ijk}\pi_{jk}^{\ \ a}=\epsilon^{ab}e_b^i.
\end{equation}
In terms of these variables, the total Hamiltonian obtained from \eqref{3D_action} takes the form
\begin{equation}\label{Hamiltonian0}
H^{(gr)}_{\text{tot}}=\frac{1}{\kappa\lambda}\!\int_{\Sigma_t}\!\!\!\!d^2x\big(\Lambda^i_t\mathcal{G}^{(gr)}_i-e^i_{\tau}\mathcal{C}_i\big)
=\frac{1}{\kappa\lambda}\!\int_{\Sigma_t}\!\!\!\!d^2x\big(\Lambda^i_t\mathcal{G}^{(gr)}_i+N^a\mathcal{V}_a^{(gr)}+N\mathcal{H}^{(gr)}\big)\,,
\end{equation}
where the curvature constraint $\mathcal{C}_i=\frac{1}{2}\epsilon^{ab}F^i_{ab}=\frac{1}{2}\epsilon^{ab}\big(2\partial_{[a}A^i_{b]}+\epsilon^{ijk}A^j_aA^k_b\big)$, contracted with Lagrange multiplier $-e^i_{\tau}$, has been naturally decomposed into two terms $N^a\mathcal{V}_a^{(gr)}+N\mathcal{H}^{(gr)}$, splitting the constraint surface in a way that reveals explicitly invariance under the spatial and time diffeomorphism transformations. The remaining term,
\begin{equation}\label{Gauss}
G^{(gr)}[\Lambda^i_t]=\frac{1}{2\kappa\lambda}\!\int_{\Sigma_t}\!\!\!\!d^2x\,\Lambda^i_t\mathcal{G}^{(gr)}_i
=-\frac{1}{\kappa\lambda}\!\int_{\Sigma_t}\!\!\!\!d^2x\,\Lambda^i_t\,D_aE^a_i
=-\frac{1}{\kappa\lambda}\!\int_{\Sigma_t}\!\!\!\!d^2x\,\Lambda^i_t\big(\partial_aE^a_i+\epsilon_{ijk}A^j_aE^a_k\big)
\end{equation}
is the SU$(2)$ Gauss constraint, with $\Lambda^i_t=\frac{1}{2}\epsilon^{ijk}\Gamma^{jk}_{\ \ \,t}$ being the Lagrange multiplier. The direct calculation \cite{Thiemann:1997ru} shows that $E_i^a$ transforms as an SU$(2)$ vector and $A^i_a$ transforms as an SU$(2)$ connection.

In analogy to four-dimensional LQG, the second object on the right hand side of \eqref{Hamiltonian0},
\begin{equation}\label{diffeomorphism}
\mathcal{V}_a^{(gr)}=-e^i_a\mathcal{C}_i=E^b_iF^i_{ba}\,,
\end{equation}
will be called the diffeomorphism constraint density. Finally the last term,
\begin{equation}\label{scalar}
\mathcal{H}^{(gr)}=\mathcal{C}_3=\frac{1}{2}\epsilon^{ab}F^3_{ab}
=\frac{1}{2}\sqrt{q}\tilde{\epsilon}^{ab}F^3_{ab},
\end{equation}
which describes dynamics of the theory on the Gauss and diffeomorphism invariant Hilbert space, will be called scalar constraint density. Here, we used the definition of the Levi-Civita tensor $\tilde{\epsilon}^{ab}=\frac{1}{\sqrt{q}}\epsilon^{ab}$ and the symbol $q$ has been introduced to denote the determinant of $q_{ab}$.

Summarizing, we recover the three-dimensional analog of the structure of LQG with the fully constrained, topological system described by the Hamiltonian \eqref{Hamiltonian0}. The Gauss constraint ${G}^{(gr)}$ \eqref{Gauss} defines the three-dimensional SU$(2)$-invariant subspace, the diffeomorphism constraint ${V}^{(gr)}$ \eqref{diffeomorphism} restricts it to the one-dimensional diffeomorphism-invariant kinematical subspace, while the scalar constraint ${H}^{(gr)}$ \eqref{scalar} implements dynamics.

\subsection{Three-dimensional Loop Quantum Gravity}\label{II.2}
\noindent
The phase space of four-dimensional LQG is spanned by holonomies of the Ashtekar-Barbero connection and fluxes of densitized vierbeins \cite{Thiemann:2007zz}. In the three-dimensional case, one can define analogical variables, smearing holonomies along some curves $\gamma$ (on a plane),
\begin{equation}
h_{\gamma}\!:=\mathcal{P}\exp\!\left(i\!\int_{\gamma}\!A^j_a(\gamma(s))\tau^j\dot{\gamma}^a(s)\right).
\end{equation}
Constructing the flux of a densitized dreibein is less straightforward. One cannot simply smear $E_i^a$ across the two-surface perpendicular to the direction $x^a$, since the perpendicular subspace is one-dimensional. However, due to the fact that the internal space is three-dimensional, it is convenient to fix the plane spanned by the two internal directions, say $1$ and $2$, with the two-dimensional coordinate space --- in other words, we choose the internal direction $3$ to be parallel to the time direction. Then we can easy define the flux across some surface $S$ to be explicitly 
\begin{equation}
E(S)\!:=\epsilon^{jkl}\!\!\int_S\!\!n_aE^a_jdx_k\wedge dx_l\,.
\end{equation}

The kinematical Hilbert space of the theory is constructed in the same way as for the four-dimensional LQG. It is a direct sum of the space of cylindrical functions of connections along each graph $\Gamma$,
\begin{equation}
\mathcal{H}_{kin}^{(gr)}:=\bigoplus_{\Gamma}\mathcal{H}_{\Gamma}^{(gr)}=L_2\big(\mathcal{A},d\mu_{AL}\big),
\end{equation}
where $\mathcal{A}$ is the space of connections and $d\mu_{AL}$ denotes the Ashtekar-Lewandowski measure \cite{Ashtekar:1994mh}. The states are defined as the cylindrical functions of all links $l_i\in\Gamma$, precisely
$\Psi_{\Gamma, f}(A):=\left<A\middle|\Gamma,f\right>:=f\big( h_{l_1}(A), h_{l_2}(A), ... , h_{l_L}(A) \big)$, with $f$ being some continuous function
$f:\text{SU}(2)^L\longrightarrow\mathds{C}$.

The basis states are individuated by fixing a graph $\Gamma$, which carries irreducible representations $D^{j_l}(h_l)$ (Wigner matrices) of spin $j$ assigned to each link $l$, and  intertwiner numbers $i_v$ that implement SU$(2)$ invariance at each node $v$. These are called spin network states and are given by the formula
\begin{equation}
\Psi_{\Gamma,j_l,i_v}(h)=\left<h\middle|\{\Gamma,{j_l},{i_v}\}\right>=\prod_{v\in\Gamma}i_v\cdot\prod_lD^{j_l}(h_l),
\end{equation}
where the product $\prod_l$ extends over all the links $l$ emanating from the node $v$, while the symbol $\cdot$ denotes contraction of the $\mathfrak{su}(2)$ indices.

Notice then that since the three-dimensional theory has no dynamical degrees of freedom, one can define any frame in the spacial hypersurface, and once fixed this will remain the same forever. In what follows, the most convenient choice is the flat Cartesian frame on the two-dimensional space, with directions $x$ and $y$ being everywhere perpendicular to each other. Moreover, since the space has been identified with the internal subspace spanned by the vectors along the directions $1$ and $2$, one can associate these directions with $x$ and $y$, respectively. Such system suggests to work with a reduced model of three-dimensional LQG, namely QRLG (truncated to a two-dimensional space). The topological property of the space, without any additional conditions, allows to implement the restriction on the spatial metric tensor to be diagonal (similarly dreibeins can be chosen to be diagonal) along some fiducial directions identified with coordinates $x,y,t$. This feature of the lower-dimensional gravity brings a remarkable simplification to LQG, which with a proper choice of the reference frame, becomes replaced with QRLG\footnote{Although the three-dimensional LQG can be rigorously simplified to QRLG, the Hamiltonian constraint operator, already solved in the four-dimensional case in \cite{Alesci:2014uha} and in the three-dimensional case in the next section, does not describe the full model. Notice that the main results in QRLG have been derived imposing homogeneity on the theory. Here we also follow this restriction, hence as long as someone considers homogeneous three-dimensional LQG, one can directly apply the results presented in this article.}.

In the case of QRLG, the basis states in the reduced kinematical Hilbert space $^{R\!}\mathcal{H}_{kin}^{(gr)}$ are constructed by projecting SU$(2)$ Wigner matrices, $D^{j_l}(h_l)$ onto coherent Livine-Speziale states $|m_l,\vec{u}_l\rangle$ \cite{Livine:2007vk} with the following graphical representations:
\begin{equation}
\raisebox{-1.5ex}{
\begin{tikzpicture}
\draw[l] (-2,0) --  node[v] {$\ \ j_l$} (-0.7,0);
\draw[l] (-0.7,-0.1) -- (-0.7,0.1);
\draw[l] (-0.5,-0.1) -- (-0.5,0.1);
\draw[l] (-0.3,0) -- (-0.5,0);
\draw[l] (0,0) circle(3mm); \node at (0.02,0) {$h_l$};
\draw[l] (0.3,0) -- (0.5,0);
\draw[l] (0.5,-0.1) -- (0.5,0.1);
\draw[l] (0.7,-0.1) -- (0.7,0.1);
\draw[l] (0.7,0) --  node[v] {$\ \ j_l$} (2,0);
\node at (-1.82,0.12) {$m$};
\node at (1.92,0.17) {$m'$};
\end{tikzpicture}
}
\!\!\!\!=
\left<j_l,m\middle|m'',\vec{u}_l\right>\left<m'',\vec{u}_l\middle|D^{j_l}(h_l)\middle|m'',\vec{u}_l\right>\left<m'',\vec{u}_l\middle|j_l,m'\right>.
\end{equation}
where $\left<j_l,i_v\middle|m_l,\vec{u}_l\right>$ denotes the reduced U$(1)$ intertwiners.

Finally the canonical, reduced variables $^{R\!}h_{l^i}$ and $^{R\!}E(S)$ are obtained by smearing diagonal holonomies and densitized dreibeins, respectively, along links $l^1$ or $l^2$ of square-graphs $\Gamma$ and across surfaces $S$ --- defined with respect to the directions in the internal space --- perpendicular to these links, respectively. Since at the quantum level we will use only these reduced variables, from now on we neglect the left uppercase symbol $^{R}$.

It is convenient now to calculate the expression for some real, positive power of the area operator $\hat{\mathbf{A}}$, which is the quantum version of the classical expression $\mathbf{A}=\int\!d^2x\sqrt{q}$, where the determinant of the spatial metric can be written as
\begin{equation}\label{determinant}
q=\frac{1}{2}\epsilon^{ab}\epsilon^{cd}q_{ac}q_{bd}=\frac{1}{2}q_{ab}E^a_iE^b_i
=\sqrt{\frac{1}{2}\tilde{\epsilon}_{ab}\tilde{\epsilon}_{cd}E^a_iE^c_iE^b_jE^d_j}\,.
\end{equation}

The action of the area operator can be defined using reduced fluxes of zweibeins and its expression is calculated analogously as the volume operator in three-dimensional LQG \cite{Alesci:2013xd}. However, since the space is only two-dimensional, we should consider smearing fluxes in the internal three-dimensional space. Then, taking into account the area of the segment $S(v_{l^i\!,\,l^j})$ containing only one divalent node $v$ in the center, from which two links $l^i$, $l^j$ emanate, we find the action of area operator $\hat{\mathbf{A}}$,
\begin{equation}\label{area2}
\begin{split}
\hat{\mathbf{A}}\big(S(v_{l^i\!,\,l^j})\big)
\ket{v_{l^i\!,\,l^j};j_i,j_j,i_{v_{l^i,l^j}}\!}_{\!R}
&=
\!\int\!\!d^2x
\bigg(\bigg|
\frac{1}{2}\tilde{\epsilon}_{kl}\tilde{\epsilon}_{kl}\hat{E}_{(k)}(S^k)\hat{E}_{(k)}(S^k)\hat{E}_{(l)}(S^l)\hat{E}_{(l)}(S^l)
\bigg|\bigg)^{\!\!\frac{1}{4}}
\ket{v_{l^i\!,\,l^j};j_i,j_j,i_{v_{l^i,l^j}}\!}_{\!R}
=\\
&=
\!\int\!\!d^2x
\Big(\Big|
\hat{E}_{(i)}(S^i)\hat{E}_{(j)}(S^j)
\Big|\Big)^{\!\!\frac{1}{2}}
\ket{v_{l^i\!,\,l^j};j_i,j_j,i_{v_{l^i,l^j}}\!}_{\!R}
=\\
&=
\kbar
\Big(\Big|
j^{(i)}\,\sigma\big(l^i,S^i\big)\,j^{(j)}\,\sigma\big(l^j,S^j\big)
\Big|\Big)^{\!\!\frac{1}{2}}
\ket{v_{l^i\!,\,l^j};j_i,j_j,i_{v_{l^i,l^j}}\!}_{\!R},
\end{split}
\end{equation}
where $\sigma\big(l^i,S^i\big)=\pm1$ depends on the relative orientation of the link $l^i$ and the surface $S^i$.

In the case of a generic node, \textit{i.e.} considering the area $S(v_{l^1\!,\,l^2})$ of a basic segment containing one tetravalent node $v_{x,y}=v_{1,2}$ (being a part of the graph $\Gamma$), one obtains
\begin{equation}\label{area4}
\hat{\mathbf{A}}^n(v_{x,y})
\ket{\Gamma;j_l,i_v}_{\!R}
=
\Bigg(\!
\kbar^2
\Bigg|\frac{j_{x-1,y}^{(1)}\!+j_{x,y}^{(1)}}{2}\ \frac{j_{x,y-1}^{(2)}\!+j_{x,y}^{(2)}}{2}\Bigg|
\Bigg)^{\!\!\!\frac{n}{2}}\!
\ket{\Gamma;j_l,i_v}_{\!R}
=
\mathbf{A}^n_{v_{x\!,y}}
\ket{\Gamma;j_l,i_v}_{\!R}\,,
\end{equation}
where
$\mathbf{A}_{v}:=\kbar\big(\big|\Sigma_{v}^{(1)}\,\Sigma_{v}^{(2)}\big|\big)^{\!\frac{1}{2}}$
denotes the eigenvalue of the area operator $\hat{\mathbf{A}}(v)$. In \eqref{area4} the object $\Sigma^{(i)}_{v}:=\frac{1}{2}\big(j^{(i)}_{v}+j^{(i)}_{v-\vec{e}_i}\big)$ represents the averaged value of the spins attached to the collinear pair of links (ingoing and outgoing) emanated from the node $v$. Specifically, the vector $\vec{e}_{i}$ denotes a unit vector along the direction $i$, such that $j^{(i)}_{v-\vec{e}_i}$ is the spin number attached to the link $l^p$ ending in $v$, with $p$ having orientation along the fiducial direction $i$. It is worth mentioning that the spectrum of $\hat{\mathbf{A}}(v)$ is discrete having minimum in the case of the fundamental representation of $\mathfrak{su}(2)$, providing the relation $\Sigma_{v}^{(1)}=\Sigma_{v}^{(2)}=1/2$. Since any lower value of area has no physical meaning, we should introduce the cutoff $\varepsilon_0$ on the regulator, which coincides with the square root of the minimal area, precisely $\varepsilon_0=\sqrt{\kbar}/2$. Notice that $\varepsilon_0$ should be understood as a lower limit for any length in the theory, which also fixes the lower value for the fiducial length, $l_0\ge\sqrt{\kbar}/2$.

\subsection{Quantum scalar field}\label{II.3}
\noindent
We quantize scalar fields minimally coupled to gravity in three-dimensions using the same representation introduced in \cite{Bilski:2015dra}. The action for the scalar is defined as
\begin{equation}\label{action}
S^{(\phi)}=\!\int_M\!\!\!\!d^{2}x\sqrt{|g|}\big(g^{\mu\nu}(\partial_{\mu}\phi)(\partial_{\nu}\phi)-V(\phi)\big),
\end{equation}
where the coupling constant has been moved to the gravitational part of the action.

The Legendre transform results in the following total Hamiltonian:
\begin{equation}\label{Hamiltonian}
H^{(\phi)}_{\text{tot}}=\!\int_{\Sigma_{(\tau)}}\!\!\!\!\!\!\!\!d^2x
\bigg(
2N^a\pi^{(\phi)}\partial_a\phi
+N\frac{1}{\sqrt{q}}
\Big(
\big(\pi^{(\phi)}\big)^{\!2}+E_i^aE_i^b\partial_a\phi\,\partial_b\phi+qV(\phi)
\Big)
\bigg)
=\!\int_{\Sigma_{(\tau)}}\!\!\!\!\!\!\!\!d^2x\big(N^a\mathcal{V}_a^{(\phi)}+N\mathcal{H}^{(\phi)}\big),
\end{equation}
where the Lagrange multipliers $N$ and $N^a$ are the lapse function and the shift vector, respectively, while $\mathcal{V}_a^{(\phi)}$ and $\mathcal{H}^{(\phi)}$ are the scalar field contributions to the vector and scalar constraints, respectively. The object denoted by $\pi^{(\phi)}$ is the momentum conjugated to $\phi$. It is worth mentioning that the term contracting derivatives of the scalar field has been using the following identity
\begin{equation}\label{inversedeterminant}
q^{ab}=\frac{1}{q}E_i^aE_i^b,
\end{equation}
which is a consequence of \eqref{determinant}.

Following the procedure introduced in \eqref{II.1}, we define the total Hamiltonian of both the gravitational and the scalar field,
\begin{equation}\label{totalHamiltonian}
H_{\text{tot}}=H^{(gr)}_{\text{tot}}+H^{(\phi)}_{\text{tot}}
=G^{(gr)}[\Lambda^i_t]+V[N^a]+H[N],
\end{equation}
where the total vector constraint, which generates diffeomorphism transformations, reads
\begin{equation}\label{vectorconstraint}
V[N^a]=\!\int_{\Sigma_{(\tau)}}\!\!\!d^2x\,N^a\bigg(\frac{1}{\kappa\lambda}\mathcal{V}_a^{(gr)}+\mathcal{V}_a^{(\phi)}\bigg).
\end{equation}
The last term in \eqref{totalHamiltonian} --- the scalar constraint--- defines the dynamics in the gauge and diffeomorphism invariant phase-space and reads
\begin{equation}\label{scalarconstraint}
H[N]
=\!\int_{\Sigma_{(\tau)}}\!\!\!\!\!\!\!\!d^2x\,N\bigg(\frac{1}{\kappa\lambda}\mathcal{H}^{(gr)}+\mathcal{H}^{(\phi)}\bigg)
=H^{(gr)}+H^{(\phi)}_{kin}\!+H^{(\phi)}_{der}\!+H^{(\phi)}_{pot}\,.
\end{equation}
It is worth noting that in the last equality we split the matter field contribution into three elements: the kinetic, derivative and potential ones, according to the sum in \eqref{Hamiltonian}.

To quantize the system with the gravitational and scalar fields we deploy the procedure introduced in \cite{Thiemann:1997rt,Thiemann:1997rq} for LQG, adapting the Polymer representation proposed in \cite{Ashtekar:2002sn,Ashtekar:2002vh,Kaminski:2005nc,Kaminski:2006ta} and later applied to the four-dimensional QRLG \cite{Bilski:2015dra} in the momentum polarization. Here, we repeat for completeness the definition of the kinematical Hilbert space, applying it to the three-dimensional model. We the obtain
\begin{equation}
\mathcal{H}_{kin}^{(\phi)}:=\overline{\big\{a_1U_{\pi_1}+...+a_nU_{\pi_n}\!:\ a_i\in\mathds{C},\, n\in\mathds{N},\, \pi_i\in\mathds{R}\big\}}=L_2\left(\bar{\mathds{R}}_{\text{Bohr}}{}^\Sigma\right),
\end{equation}
where $\pi:\Sigma\rightarrow\mathbb{R}$ is  the function with finite support $supp\, \pi=\{v_1,\ldots,v_n\}$. The discretized scalar field $\phi_v=\{\phi(v_1),\ldots,\phi(v_n)\}$, defined at the nodes $\{v_1,\ldots,v_n\}\in\Gamma$, represents the state
\begin{equation}
U_\pi(\phi):=\langle \phi| U_{\{v_1,..,v_n\},\{\pi_{v_1},\pi_{v_n}\}}\rangle=e^{i\sum_{v\in\Sigma}\pi^{(\phi)}_v \phi_v}\,.
\end{equation}
In order to introduce a scalar product, we can introduce the Bohr measure,
\begin{equation}
\int_{\bar{\mathds{R}}_{\text{Bohr}}}\!\!\!\!\!\!\!\!d\mu_{\text{Bohr}}(\phi)e^{i\pi_x\phi_x}=\delta_{0,x}\,,
\end{equation}
In which $\bar{\mathds{R}}_{\text{Bohr}}$ denotes the Bohr compactification of a real line.
Hence the field can be viewed as a set of square-integrable functions in $\mathcal{H}_{kin}^{(\phi)}$. The scalar product defined in this way on the set of notes $\{v_1,\ldots,v_n\}\cup\{v'_1,\ldots,v'_m\}$ satisfies the orthogonality relation
\begin{equation}
\bracket{U_{\pi}}{U_{\pi'}}:=\delta_{\pi,\pi'}\,.
\end{equation}

Finally, the basic operators act as follows:
\begin{equation}
\begin{split}\label{canonicaloperators}
\langle{\phi}|\hat{U}_\pi\ket{U_{\pi'}}
&=
e^{i\sum_{v\in\Sigma}\pi^{(\phi)}_v\hat{\phi}_v}\ket{U_{\pi'}}
=\langle{\phi}\ket{U_{\pi+\pi'}}
=e^{i\left(\sum_{v\in\Sigma}\pi^{(\phi)}_v \phi_v+ \sum_{v'\in\Sigma}{\pi'_{v'}}^{\!\!\!\!(\phi)}\phi_{v'}\right)}
,\\
\hat{\Pi}(V)\ket{U_{\pi}}
&=
-i\hbar\sum_{v\in V}\frac{\partial}{\partial\phi(v)}\ket{U_{\pi}}
=\hbar\sum_{v\in V}\pi_v^{(\phi)}\ket{U_{\pi}},
\end{split}
\end{equation}
with $\Pi(V)$ representing the scalar field momentum smeared over volume $V\subseteq\Sigma$ and $v\in V$ being the subset of points $v\in \Sigma$. Next, for each point we define a segment of radius $\varepsilon$ that allows us to smear the momentum in the following way,
\begin{equation}
\Pi^{(\phi)}(v):=\frac{1}{\varepsilon}\int\!\!d^2u\, \chi_{\varepsilon}(v,u)\pi^{(\phi)}(u).
\end{equation}
The new introduced object, $\chi_{\varepsilon}(x,y)$, is the characteristic function defined as
\begin{equation}\label{characteristic}
f(x)=\int\!\!d^2y\,\delta^2(x-y)f(y)=\lim_{\varepsilon\to0}\frac{1}{\varepsilon^2}\!\int\!\!d^2y\,f(y)\chi_{\varepsilon}(x,y),
\end{equation}
which in the limit $\varepsilon\to0$ behaves as a Dirac delta. Taking this limit together with increasing the number of points to infinity, we recover the classical limit.

In the case of the planar QRLG, the states in the total Hilbert space $\mathcal{H}_{kin}$ are described in the following way:
\begin{equation}
\ket{\Gamma;j_l,i_v;U_{\pi}}_{\!R}=\!
\ket{\Gamma;j_l,i_v}_{\!R}\otimes\ket{\Gamma;U_{\pi}}_{\!R}
=\!\vket{
\begin{tikzpicture}
\draw[l] (-1.2,0) --  node[v] {$j_{x-1,y}^{(1)}$} (0,0);
\draw[l] (-1.2,-0.1) -- (-1.2,0.1);
\draw[l] (-1.4,-0.1) -- (-1.4,0.1);
\draw[l] (-1.4,0) -- (-1.7,0);
\draw[l] (-1.85,0) circle(1.5mm); \node at (-1.8,0.4) {$h_{^{x\texttt{-\!}1\!,y}}^{\!(1)}$};
\draw[l] (-2.0,0) -- (-2.3,0);
\draw[l] (-2.3,-0.1) -- (-2.3,0.1);
\draw[l] (-2.5,-0.1) -- (-2.5,0.1);
\draw[l] (-3.7,0) -- node[v] {$j_{x-1,y}^{(1)}$} (-2.5,0);
\draw[l] (-3.7,-0.8) -- (-3.7,0.8);
\draw[l] (-3.7,0) -- (-3.9,0);
\draw[l] (0,0) --  node[v] {$j_{x,y}^{(1)}$} (1.5,0);
\draw[l] (1.2,-0.1) -- (1.2,0.1);
\draw[l] (1.4,-0.1) -- (1.4,0.1);
\draw[l] (1.4,0) -- (1.7,0);
\draw[l] (1.85,0) circle(1.5mm); \node at (1.9,0.4) {$h_{^{x\!,y}}^{\!(1)}$};
\draw[l] (2.0,0) -- (2.3,0);
\draw[l] (2.3,-0.1) -- (2.3,0.1);
\draw[l] (2.5,-0.1) -- (2.5,0.1);
\draw[l] (2.5,0) -- node[v] {$j_{x,y}^{(1)}$} (3.7,0);
\draw[l] (3.7,-0.8) -- (3.7,0.8);
\draw[l] (3.7,0) -- (3.9,0);
\draw[l] (0,0.5) --  node[v] {$j_{x,y}^{(2)}$} (0,1.2);
\draw[l] (0,0) -- (0,0.5);
\draw[l] (-0.1,1.2) -- (0.1,1.2);
\draw[l] (-0.1,1.4) -- (0.1,1.4);
\draw[l] (0,1.4) -- (0,1.7);
\draw[l] (0,1.85) circle(1.5mm); \node at (-0.5,1.8) {$h_{^{x\!,y}}^{\!(2)}$};
\draw[l] (0,2.0) -- (0,2.3);
\draw[l] (-0.1,2.3) -- (0.1,2.3);
\draw[l] (-0.1,2.5) -- (0.1,2.5);
\draw[l] (0,2.5) --  node[v] {$j_{x,y}^{(2)}$} (0,3.7);
\draw[l] (0,3.7) -- (0,3.9);
\draw[l] (-0.8,3.7) -- (0.8,3.7);
\draw[l] (0,-1.2) --  node[v] {$j_{x,y-1}^{(2)}$} (0,-0.5);
\draw[l] (0,-0.5) -- (0,0);
\draw[l] (-0.1,-1.2) -- (0.1,-1.2);
\draw[l] (-0.1,-1.4) -- (0.1,-1.4);
\draw[l] (0,-1.4) -- (0,-1.7);
\draw[l] (0,-1.85) circle(1.5mm); \node at (-0.55,-1.95) {$h_{^{x\!,y\texttt{-\!}1}}^{\!(2)}$};
\draw[l] (0,-2.0) -- (0,-2.3);
\draw[l] (-0.1,-2.3) -- (0.1,-2.3);
\draw[l] (-0.1,-2.5) -- (0.1,-2.5);
\draw[l] (0,-3.7) --  node[v] {$j_{x,y-1}^{(2)}$} (0,-2.5);
\draw[l] (0,-3.9) -- (0,-3.7);
\draw[l] (0.8,-3.7) --  (-0.8,-3.7);
\node at (-2.8,-0.25) {$e^{i\pi_{x\texttt{-\!}1\!,y}\phi_{x\texttt{-\!}1\!,y}}$};
\node at (0.7,-0.25) {$e^{i\pi_{x\!,y}\phi_{x\!,y}}$};
\node at (3.59,-0.25) {$e^{i\pi_{x\texttt{+\!}1\!,y}\phi_{x\texttt{+\!}1\!,y}}$};
\node at (0.9,3.45) {$e^{i\pi_{x\!,y\texttt{+\!}1}\phi_{x\!,y\texttt{+\!}1}}$};
\node at (0.9,-3.45) {$e^{i\pi_{x\!,y\texttt{-\!}1}\phi_{x\!,y\texttt{-\!}1}}$};
\end{tikzpicture}\!\!\!\!\!\!
}.\!
\label{4gr_sc_graph}
\end{equation}
The scalar field state is recovered by assigning at the each node $v_{p,q}\!\in\!\Gamma$ (placed at $(x,y)$) the point holonomy $e^{i\pi_{p,q}\phi_{p,q}}$ with the real coefficient $\pi_{p,q}$, while the gravity is described by the spin numbers $j_{p,q}^{(i)}$ at the associated links $l_{p,q}^{(i)}$ and the reduced intertwiners $i_v$ at nodes. For a detailed explanation of this graphical representation, we refer to \cite{Bilski:2015dra}.

\section{Quantization}\label{III}

In this section, we address the quantization of the model we defined above, dealing first with the lattice regularization of the scalar constraint \eqref{III.1}, and then in section \eqref{III.2} with its quantization, hence providing an analytical expression for the Hamiltonian operator. The Hamiltonian operator we find clearly coincides with the Hamiltonian operator of $2+1$ dimensional LQC.
		
\subsection{Regularization}\label{III.1}
\noindent
In order to quantize the scalar constraint using LQG techniques, we perform the lattice regularization of the classical canonical variables. This is achieved by recasting expression \eqref{scalarconstraint} in terms of the holonomies and fluxes for both the gravitational and the matter components.

The gravitational term is regularized using the discretization procedure of spacetime introduced by Thiemann \cite{Thiemann:1996aw}, further restricted to the square-graphs. Using first the method explained in the appendix \eqref{AppendixA}, one can write the following regularized expressions for the all the terms that contribute to the Hamiltonian constraint:
\begin{equation}
\begin{split}
H^{(gr)}
&=
\frac{1}{2\kappa\lambda}\!\int_{\Sigma_{(\tau)}}\!\!\!\!\!\!\!\!d^2x\,N
\sqrt{q}\tilde{\epsilon}^{ab}F^3_{ab}
=
\frac{1}{2\kappa\lambda}\lim_{\varepsilon\to0}\frac{1}{\varepsilon^2}\!\int_{\Sigma_{(\tau)}}\!\!\!\!\!\!\!\!d^2x\!\int_{\Sigma_{(\tau)}}\!\!\!\!\!\!\!\!d^2y\,N(x)
\bigg(\frac{1}{2}\tilde{\epsilon}_{cd}\tilde{\epsilon}_{ef}E^c_i(x)E^d_j(x)E^e_i(x)E^f_j(x)\!\bigg)^{\!\!\frac{1}{4}}
\tilde{\epsilon}^{ab}F^3_{ab}(y)\,\chi_{\varepsilon}(x,y),
\label{gravitational1}
\end{split}
\end{equation}
\begin{equation}
\begin{split}
H^{(\phi)}_{kin}
&=
\!\int_{\Sigma_{(\tau)}}\!\!\!\!\!\!\!\!d^2x\,N\frac{\big(\pi^{(\phi)}\big)^2}{\sqrt{q}}
=\\
&=
\lim_{\varepsilon\to0}
\!\int_{\Sigma_{(\tau)}}\!\!\!\!\!\!\!\!d^2w\frac{\pi^{(\phi)}(w)\,\chi_{\varepsilon}(w,x)}{\varepsilon}
\!\int_{\Sigma_{(\tau)}}\!\!\!\!\!\!\!\!d^2x\frac{\pi^{(\phi)}(x)\,\chi_{\varepsilon}(x,y)}{\varepsilon}
\!\int_{\Sigma_{(\tau)}}\!\!\!\!\!\!\!\!d^2y
\!\int_{\Sigma_{(\tau)}}\!\!\!\!\!\!\!\!d^2z
\,N(y)\,\mathbf{A}^{\!-\frac{1}{2}}(y,\varepsilon)\,\mathbf{A}^{\!-\frac{1}{2}}(z,\varepsilon)
\,\chi_{\varepsilon}(y,z)
=\\
&=
\frac{2^{15}}{3^4(\gamma\kappa)^4}\lim_{\varepsilon\to0}
\!\int_{\Sigma_{(\tau)}}\!\!\!\!\!\!\!\!d^2w\frac{\pi^{(\phi)}(w)\,\chi_{\varepsilon}(w,x)}{\varepsilon}
\!\int_{\Sigma_{(\tau)}}\!\!\!\!\!\!\!\!d^2x\frac{\pi^{(\phi)}(x)\,\chi_{\varepsilon}(x,y)}{\varepsilon}
\!\int_{\Sigma_{(\tau)}}\!\!\!\!\!\!\!\!d^2y
\!\int_{\Sigma_{(\tau)}}\!\!\!\!\!\!\!\!d^2z
\,N(y)\,
\\
&\times
\tilde{\epsilon}^{ac}\tilde{\epsilon}^{bd}
\Big\{A_{a}^i,\mathbf{A}^{\!\frac{3}{4}}(y,\varepsilon)\Big\}\Big\{A_{b}^i,\mathbf{A}^{\!\frac{3}{4}}(y,\varepsilon)\Big\}
\Big\{A_{c}^j,\mathbf{A}^{\!\frac{3}{4}}(z,\varepsilon)\Big\}\Big\{A_{d}^j,\mathbf{A}^{\!\frac{3}{4}}(z,\varepsilon)\Big\}
\,\chi_{\varepsilon}(y,z),
\label{kinetic1}
\end{split}
\end{equation}
\begin{equation}
\begin{split}
H^{(\phi)}_{der}
&=
\!\int_{\Sigma_{(\tau)}}\!\!\!\!\!\!\!\!d^2x\,N\frac{1}{\sqrt{q}}E^a_iE^b_i\partial_a\phi\,\partial_b\phi
=
\!\int_{\Sigma_{(\tau)}}\!\!\!\!\!\!\!\!d^2x\,N
\partial_a\phi(x)\frac{E^a_i}{\mathbf{A}^{\!\frac{1}{2}}(x,\varepsilon)}\,\partial_b\phi(y)\frac{E^b_i}{\mathbf{A}^{\!\frac{1}{2}}(y,\varepsilon)}
\,\chi_{\varepsilon}(x,y)
=\\
&=
\frac{2^4}{(\gamma\kappa)^2}\lim_{\varepsilon\to0}\!\int_{\Sigma_{(\tau)}}\!\!\!\!\!\!\!\!d^2x\!\int_{\Sigma_{(\tau)}}\!\!\!\!\!\!\!\!d^2y\,N(x)
\,\tilde{\epsilon}^{ac}\,\tilde{\epsilon}^{bd}
\partial_a\phi(x)\Big\{A_c^i,\mathbf{A}(x,\varepsilon)\Big\}
\partial_b\phi(y)\Big\{A_d^i,\mathbf{A}(y,\varepsilon)\Big\}
\,\chi_{\varepsilon}(x,y)
\label{derivative1}
\end{split}
\end{equation}
and
\begin{equation}
\begin{split}
H^{(\phi)}_{pot}
&=\!\int_{\Sigma_{(\tau)}}\!\!\!\!\!\!\!\!d^2x\,N\sqrt{q}\,V(\phi)
=
\!\int_{\Sigma_{(\tau)}}\!\!\!\!\!\!\!\!d^2x\,N(x)
\bigg(\frac{1}{2}\tilde{\epsilon}_{ab}\tilde{\epsilon}_{cd}E^a_i(x)E^b_j(x)E^c_i(x)E^d_j(x)\!\bigg)^{\!\!\frac{1}{4}}
\,V\big(\phi(x)\big),
\label{potential1}
\end{split}
\end{equation}
where we used the definition of the characteristic function $\chi_{\varepsilon}(x,y)$ in \eqref{characteristic}, and the definition of the coordinate area of surface $S_{\varepsilon}(x)$,
\begin{equation}\label{areadecomposition}
\mathbf{A}\big(S_{\varepsilon}(x)\big):=\mathbf{A}(x,\varepsilon)=\varepsilon^2\sqrt{q}(x)+\mathcal{O}(\varepsilon^3).
\end{equation}
It is worth noting that the expressions \eqref{gravitational1}, \eqref{kinetic1}, \eqref{derivative1} and \eqref{potential1} are independent on the regularization parameter $\varepsilon$ at the leading order, providing a finite outcome in the limit $\varepsilon\to0$. Moreover, the regularization procedure applied to formulas \eqref{kinetic1} and \eqref{derivative1} removed the gravitational degrees of freedom from denominators.

To represent the four contributions to the scalar constraint written above in the form of a figure as in \eqref{4gr_sc_graph}, we follow the Thiemann's procedure. The latter was first developed in order to triangulate the spatial manifold in presence of pure gravity \cite{Thiemann:1996aw} and then applied to the model endowed with matter fields \cite{Thiemann:1997rt}. This procedure has been successfully adapted to cuboidal graphs in QRLG \cite{Bilski:2015dra,Bilski:2016pib} and finally simplified to the case of homogeneous models \cite{Bilski:2017ijd}. Therefore we can simply replace the integration over the spatial hypersurface $\lim_{\varepsilon\to0}\int_{\Sigma}d^2x$ with the sum $\sum_{v\in\Gamma}l_0^2$ over all the nodes $v$. This should be also modified by factor 2 for each contraction of orientation-dependent operators --- after summing over positive and negative orientations of links we obtain the standard contraction of directions.

Notice that the simplified method for homogeneous systems \cite{Bilski:2017ijd} changes the standard tessellation procedure --- for square-graphs we should call it squareation --- so to replace the integral $\int_{\Sigma}$ with the sum not only over all the nodes $v\in\Gamma$, but also over all the tetrahedra $\Delta_{l,l'\!,l''}$ created by triples of links $\{l,l'\!,l''\}$ emanating from $v$ --- see \cite{Bilski:2015dra} for comparison. In that case, since each node $v$ is always surrounded by two pairs of collinear links oriented along fixed perpendicular directions, these links would create four triangles around the node and for each triangle, the remaining three would coincide with the three ``small triangles'' that one should construct to triangulate the lattice --- see \textit{e.g.} \cite{Thiemann:1997ru}. Since the square-structure of the graph is preserved by the application of the diffeomorphism constraint, the three ``small triangles'' that would correspond to each of the four pairs of neighbor links, should be replaced here by ``small squares'' supported on the remaining pairs --- one for each of the four pairs of links. Finally, the integration over each triangle, namely $\int_{\Delta_{l,l'}}$, would recast in the sum over the four possible choices of pairs of perpendicular links $\{l,l'\}$ around the node $v$. \\

Coming back to the simplified procedure and using the relation \eqref{loopexpansion}, we derive the result
\begin{equation}
\begin{split}
H^{(gr)}
&=
\frac{l_0^4}{32\pi G\lambda}\lim_{\varepsilon\to0}
\!\!\sum_{v\in\mathbf{A}(\Gamma)}\sum_{v'\in\mathbf{A}(\Gamma)\!}\!N(v)
\bigg(\bigg|
\frac{1}{2}\tilde{\epsilon}_{rs}\tilde{\epsilon}_{tu}E_i\big(S^r(v)\big)E_i\big(S^t(v)\big)E_j\big(S^s(v)\big)E_j\big(S^u(v)\big)
\bigg|\bigg)^{\!\!\frac{1}{4}}
\\
&\times
\tilde{\epsilon}^{pq}F^3_{ab}(v')\,\delta^a_{l^p\!(v)}\delta^b_{l^q\!(v)}
\delta_{v,v'}
\\
&=
-\frac{l_0^4}{8\pi G\lambda}\lim_{\varepsilon\to0}\frac{1}{\varepsilon^2}
\sum_{v}\!N(v)
\bigg(\bigg|
\frac{1}{2}\tilde{\epsilon}_{rs}\tilde{\epsilon}_{tu}E_i\big(S^r(v)\big)E_i\big(S^t(v)\big)E_j\big(S^s(v)\big)E_j\big(S^u(v)\big)
\bigg|\bigg)^{\!\!\frac{1}{4}}
\\
&\times
\tilde{\epsilon}^{pq}\,\text{tr}\big(\tau^3h_{p\circlearrowleft q}(v)\big),
\label{gravitational2}
\end{split}
\end{equation}
where in the second line we applied the Kronecker delta ($v=v'$). Analogously, for the scalar field components we apply relation \eqref{holonomyexpansion}, and obtain
\begin{equation}
\begin{split}
H^{(\phi)}_{kin}
&=
\frac{2^{15}l_0^6}{3^4(8\pi G\gamma)^4}
\lim_{\varepsilon\to0}\frac{1}{\varepsilon^4}
\sum_{v}N(v)
\,\big(\Pi^{(\phi)}\big)^{\!2}(v)
\times\\
&\times
\tilde{\epsilon}^{tu}
\,\text{tr}\Big(\tau^ih_t^{-1}(v)\Big\{h_t(v),\mathbf{A}^{\!\frac{3}{4}}(v)\Big\}\Big)
\,\text{tr}\Big(\tau^jh_u^{-1}(v)\Big\{h_u(v),\mathbf{A}^{\!\frac{3}{4}}(v)\Big\}\Big)
\times\\
&\times
\tilde{\epsilon}^{t'\!u'}
\,\text{tr}\Big(\tau^ih_{t'}^{-1}(v)\Big\{h_{t'}(v),\mathbf{A}^{\!\frac{3}{4}}(v)\Big\}\Big)
\,\text{tr}\Big(\tau^jh_{u'}^{-1}(v)\Big\{h_{u'}(v),\mathbf{A}^{\!\frac{3}{4}}(v)\Big\}\Big),
\label{kinetic2}
\end{split}
\end{equation}
\begin{equation}
\begin{split}
H^{(\phi)}_{der}
&=
\frac{2^4l_0^4}{(8\pi G\gamma)^2}
\lim_{\varepsilon\to0}\frac{1}{\varepsilon^2}
\sum_{v}N(v)
\,\tilde{\epsilon}^{pr}
\frac{e^{\rho(\phi_{v\texttt{\,+}\vec{e}_p}\!-\phi_{v})}\!-e^{\rho(\phi_{v}-\phi_{v\texttt{\,-}\vec{e}_p})}\!}{2\varepsilon\rho}
\,\text{tr}\Big(\tau^ih_r^{-1}(v)\Big\{h_r(v),\mathbf{A}(v)\Big\}\Big)
\times\\
&\times
\,\tilde{\epsilon}^{qs}
\frac{e^{\rho(\phi_{v\texttt{\,+}\vec{e}_q}\!-\phi_{v})}\!-e^{\rho(\phi_{v}-\phi_{v\texttt{\,-}\vec{e}_q})}\!}{2\varepsilon\rho}
\,\text{tr}\Big(\tau^ih_s^{-1}(v)\Big\{h_s(v),\mathbf{A}(v)\Big\}\Big),
\label{derivative2}
\end{split}
\end{equation}
\begin{equation}
\begin{split}
H^{(\phi)}_{pot}
&=
l_0^2\sum_{v}N(v)
\bigg(\bigg|
\frac{1}{2}\tilde{\epsilon}_{rs}\tilde{\epsilon}_{tu}E_i\big(S^r(v)\big)E_i\big(S^t(v)\big)E_j\big(S^s(v)\big)E_j\big(S^u(v)\big)
\bigg|\bigg)^{\!\!\frac{1}{4}}
\,V\big(\phi(v)\big),
\label{potential2}
\end{split}
\end{equation}
where we adapted the discrete expression for the derivative\footnote{Notice that the choice of an arbitrary factor $\rho$ in the exponents and in the denominator introduces a definition ambiguity. This is a generic property of the polymer quantization, which reflects the freedom in fixing a minimum scale for the scalar field discretization.} in the polymer quantization --- see \textit{e.g.} \cite{Ashtekar:2002sn,Ashtekar:2002vh,Kaminski:2005nc,Kaminski:2006ta} ---
$$
\partial_p\phi(v)\approx\frac{1}{2\varepsilon\rho}
\Big(\,\text{exp}\big(\rho(\phi_{v+\vec{e}_p}-\phi_{v})\big)-\,\text{exp}\big(\rho(\phi_{v}-\phi_{v-\vec{e}_p})\big)\Big)\,.
$$
The object $\phi_{v+\vec{e}_p}$ is the field at the point $v+\vec{e}_p$, which is the nearest node of $v$ along the link $l^p$ of length $\varepsilon$.

\subsection{Action of the scalar constraint operator}\label{III.2}
\noindent
The lattice-regularized scalar constraint is quantized by deploying the canonical procedure: holonomies, areas and dynamical matter variables are promoted into quantum operators that act on the states \eqref{4gr_sc_graph} belonging to the total kinematical Hilbert space $\mathcal{H}_{kin}$. Finally, the Poisson brackets in \eqref{kinetic2} and \eqref{derivative2} are replaced by commutators multiplied by $1/i\hbar$. In what follows, we continue to discuss separately the four terms of the Hamiltonian constrained operator we obtained hitherto, \textit{i.e.}
\begin{equation}\label{HCO}
\hat{H}\!\ket{\Gamma;j_l,i_v;U_{\pi}}_{\!R}
=\Big(\hat{H}^{(gr)}+\hat{H}^{(\phi)}_{kin}\!+\hat{H}^{(\phi)}_{der}\!+\hat{H}^{(\phi)}_{pot}\Big)\!\ket{\Gamma;j_l,i_v;U_{\pi}}_{\!R}\,.
\end{equation}

Then the quantum operators corresponding to the components of the regularized Hamiltonian \eqref{gravitational2}, \eqref{kinetic2}, \eqref{derivative2} and \eqref{potential2}, respectively, act as
\begin{equation}
\begin{split}
\hat{H}^{(gr)}
\!\ket{\Gamma;j_l,i_v;U_{\pi}}_{\!R}
&=
-\frac{\gamma\hbar}{\kbar\lambda}\lim_{\varepsilon\to0}\frac{l_0^4}{\varepsilon^2}
\sum_{v}N(v)\,\hat{\mathbf{A}}(v)
\,\epsilon^{ij}\,\text{tr}\Big(\big(\hat{h}_{i\circlearrowleft j}-\hat{h}_{i\circlearrowleft j}^{-1}\big)\tau^3\Big)
\!\ket{\Gamma;j_l,i_v;U_{\pi}}_{\!R}\Big|_{i,j=1,2},
\label{gravitational3}
\end{split}
\end{equation}
\begin{equation}
\begin{split}
\hat{H}^{(\phi)}_{kin}
\!\ket{\Gamma;j_l,i_v;U_{\pi}}_{\!R}
&=
\frac{2^{19}}{3^4\kbar^4}\lim_{\varepsilon\to0}\frac{l_0^6}{\varepsilon^4}
\sum_{v}N(v)
\,\big(\hat{\Pi}^{(\phi)}\big)^{\!2}(v)
\,\epsilon^{kl}
\,\text{tr}\Big(\tau^i\hat{h}_k^{-1}(v)\,\hat{\mathbf{A}}^{\!\frac{3}{4}}(v)\,\hat{h}_k(v)\Big)
\,\text{tr}\Big(\tau^j\hat{h}_l^{-1}(v)\,\hat{\mathbf{A}}^{\!\frac{3}{4}}(v)\,\hat{h}_l(v)\Big)
\times\\
&\times
\,\epsilon^{mn}
\,\text{tr}\Big(\tau^i\hat{h}_m^{-1}(v)\,\hat{\mathbf{A}}^{\!\frac{3}{4}}(v)\,\hat{h}_m(v)\Big)
\,\text{tr}\Big(\tau^j\hat{h}_n^{-1}(v)\,\hat{\mathbf{A}}^{\!\frac{3}{4}}(v)\,\hat{h}_n(v)\Big)
\!\ket{\Gamma;j_l,i_v;U_{\pi}}_{\!R}\Big|_{k,l,m,n=1,2},
\label{kinetic3}
\end{split}
\end{equation}
\begin{equation}
\begin{split}
\hat{H}^{(\phi)}_{der}
\!\ket{\Gamma;j_l,i_v;U_{\pi}}_{\!R}
&=
-\frac{2^6}{\kbar^2}\lim_{\varepsilon\to0}\frac{l_0^4}{\varepsilon^2}
\sum_{v}N(v)
\,\epsilon^{km}
\frac{e^{\rho(\hat{\phi}_{v\texttt{\,+}\vec{e}_k}\!-\hat{\phi}_{v})}\!-e^{\rho(\hat{\phi}_{v}-\hat{\phi}_{v\texttt{\,-}\vec{e}_k})}\!}{2\varepsilon\rho}
\,\text{tr}\Big(\tau^i\hat{h}_m^{-1}(v)\,\hat{\mathbf{A}}(v)\,\hat{h}_m(v)\Big)
\times\\
&\times
\,\epsilon^{ln}
\frac{e^{\rho(\hat{\phi}_{v\texttt{\,+}\vec{e}_l}\!-\hat{\phi}_{v})}\!-e^{\rho(\hat{\phi}_{v}-\hat{\phi}_{v\texttt{\,-}\vec{e}_l})}\!}{2\varepsilon\rho}
\,\text{tr}\Big(\tau^i\hat{h}_n^{-1}(v)\,\hat{\mathbf{A}}(v)\,\hat{h}_n(v)\Big)
\!\ket{\Gamma;j_l,i_v;U_{\pi}}_{\!R}\Big|_{k,l,m,n=1,2}
\label{derivative3}
\end{split}
\end{equation}
and
\begin{equation}
\begin{split}
\hat{H}^{(\phi)}_{pot}
\!\ket{\Gamma;j_l,i_v;U_{\pi}}_{\!R}
&=
l_0^2\sum_{v}N(v)\,\hat{\mathbf{A}}(v)\,\hat{V}\big(\phi(v)\big)
\!\ket{\Gamma;j_l,i_v;U_{\pi}}_{\!R}\,.
\label{potential3}
\end{split}
\end{equation}
Notice that we already summed over the orientations.

It is worth mentioning that the operator $\hat{V}\big(\phi(v)\big)$ in \eqref{potential3} should be properly defined as a polynomial function of field variables in terms of their ``exponential'' versions --- \textit{e.g.} like the derivative operator. However, here we are interested only in the leading order corrections coming from the discretization procedure. In other words, if we do not impose \textit{a priori} the proper weight of the potential that may lead to redensitization of the operator $\hat{V}\big(\phi(v)\big)$, any specific choice of the regularized expression will be relevant for us. For a discussion about possible choices of the functional $V\big(\phi(v)\big)$, depending on the chosen representation of the scalar field, we refer the reader to \cite{Bilski_Thesis}.

As a result of the quantization achieved, the operator \eqref{HCO} is represented by a sum of basic segments (squares) \eqref{4gr_sc_graph} that extend over all the nodes of the graph $\Gamma$. Each of the basic segments is constructed from a sum of elements acting on the central node $v$ surrounded by four nearest neighbor nodes --- and it is rescaled by a factor 1/4, that removes overlap of segments. This segmental composition allows to restrict calculations to a basic segment and to represent the final result as the sum over all the segments --- rescaled by 1/4.

In what follows we apply the formulas \eqref{BCHtracetrick}, \eqref{BCHloop} and \eqref{area4} to the actions above. It is worth noting that if one would like to consider the next to the leading order corrections or the exact expressions from \eqref{BCHtracetrick} and \eqref{BCHloop}, one should also derive the possible corrections coming from the discretized representation of the potential operator $\hat{V}\big(\phi(v)\big)$ and the derivative operator
$\big(e^{\rho(\hat{\phi}_{v\texttt{\,+}\vec{e}_p}\!-\hat{\phi}_{v})}\!-e^{\rho(\hat{\phi}_{v}-\hat{\phi}_{v\texttt{\,-}\vec{e}_p})}\big)/(2\varepsilon\rho)$.\\

Finally, we find the actions of the four contributions to the Hamiltonian constraint operator, which read 
\begin{equation}
\begin{split}
\hat{H}^{(gr)}
\!\ket{\Gamma;j_l,i_v;U_{\pi}}_{\!R}
&\approx
\frac{\gamma\hbar}{\kbar\lambda}\lim_{\varepsilon\to0}l_0^2
\sum_{v}N(v)\,\mathbf{A}_{v}
\,\epsilon^{ij}\epsilon_{ij}c_{(i)\!}(v)\,c_{(j)\!}(v)
\!\ket{\Gamma;j_l,i_v;U_{\pi}}_{\!R}\Big|_{i,j=1,2},
\label{gravitational4}
\end{split}
\end{equation}
\begin{equation}
\begin{split}
\hat{H}^{(\phi)}_{kin}
\!\ket{\Gamma;j_l,i_v;U_{\pi}}_{\!R}
&\approx
\frac{1}{2\kbar^4}\lim_{\varepsilon\to\varepsilon_0}l_0^2
\sum_{v}N(v)\frac{\mathbf{A}_v^3}{\Sigma^{(k)}_v\Sigma^{(l)}_v\Sigma^{(m)}_v\Sigma^{(n)}_v}
\,\big(\hbar\pi^{(\phi)}_v\big)^{\!2}
\times\\
&\times
\,\epsilon^{kl}
\,\epsilon^{mn}
\delta_{ik}\delta_{jl}\delta_{im}\delta_{jn}
\!\ket{\Gamma;j_l,i_v;U_{\pi}}_{\!R}\Big|_{k,l,m,n=1,2},
\label{kinetic4}
\end{split}
\end{equation}
\begin{equation}
\begin{split}
\hat{H}^{(\phi)}_{der}
\!\ket{\Gamma;j_l,i_v;U_{\pi}}_{\!R}
&\approx
\frac{1}{\kbar^2}\lim_{\varepsilon\to\varepsilon_0}l_0^2
\sum_{v}N(v)\frac{\mathbf{A}_v^2}{\Sigma^{(m)}_v\Sigma^{(n)}_v}
\,\epsilon^{km}
\frac{e^{\rho(\hat{\phi}_{v\texttt{\,+}\vec{e}_k}\!-\hat{\phi}_{v})}\!-e^{\rho(\hat{\phi}_{v}-\hat{\phi}_{v\texttt{\,-}\vec{e}_k})}\!}{2\varepsilon\rho}
\delta_{im}
\times\\
&\times
\,\epsilon^{ln}
\frac{e^{\rho(\hat{\phi}_{v\texttt{\,+}\vec{e}_l}\!-\hat{\phi}_{v})}\!-e^{\rho(\hat{\phi}_{v}-\hat{\phi}_{v\texttt{\,-}\vec{e}_l})}\!}{2\varepsilon\rho}
\delta_{in}
\!\ket{\Gamma;j_l,i_v;U_{\pi}}_{\!R}\Big|_{k,l,m,n=1,2}
\label{derivative4}
\end{split}
\end{equation}
\begin{equation}
\begin{split}
\hat{H}^{(\phi)}_{pot}
\!\ket{\Gamma;j_l,i_v;U_{\pi}}_{\!R}
&=
l_0^2\sum_{v}N(v)
\,\mathbf{A}_v\,{\scriptstyle V}\big(\phi(v)\big)
\!\ket{\Gamma;j_l,i_v;U_{\pi}}_{\!R},
\label{potential4}
\end{split}
\end{equation}
where $\big(e^{\rho(\phi_{v\texttt{\,+}\vec{e}_p}\!-\phi_{v})}\!-e^{\rho(\phi_{v}-\phi_{v\texttt{\,-}\vec{e}_p})}\big)/(2\varepsilon\rho)$ and
${\scriptstyle V}\big(\phi(v)\big)$
are respectively the eigenvalues of the derivative operator --- acting by multiplication --- and of the potential operator --- the action of which depends on the structure of the potential function. Notice that we introduced the cutoff $\varepsilon_0$ and derived the leading terms in the eigenvalues of the contributions to the Hamiltonian constraint, without using the large-$j$ expansion --- see also \cite{Bilski:2017ijd} for a discussion. This was also a necessary assumption that led to semiclassical result in the standard tessellation method \cite{Alesci:2014uha,Bilski:2015dra}.

\subsection{Continuum limit}\label{III.3}
\noindent
Finally, we can calculate the continuum limit of the leading order of matrix element of the Hamiltonian operator, defined as
$$
{}_{\raisebox{-2.5pt}{\scriptsize$R$\!}}\big<\hat{H}\big>_{\!\!R}:=
{}_{\raisebox{-1.5pt}{\scriptsize$R\!\!$}}\bra{\Gamma;j_l,i_v;\underline{n}_l,\underline{i}_v}
\hat{H}^{(gr)}+\hat{H}^{(\phi)}_{kin}\!+\hat{H}^{(\phi)}_{der}\!+\hat{H}^{(\phi)}_{pot}
\!\ket{\Gamma;j_l,i_v;\underline{n}_l,\underline{I}_v}_{\!R}\,.
$$
Taking into account the segmental structure of the square-lattice, this limit is obtained simply by reducing the distance between the nodes, \textit{i.e.} $\varepsilon\to\varepsilon_0$, while simultaneously increasing their number --- this amounts to reintroduce a continuos structure, namely $\sum_{v}l_0^2\to\int\!dx^2$. This procedure then leads to the result
\begin{equation}
\begin{split}
{}_{\raisebox{-2.5pt}{\scriptsize$R$\!}}\big<\hat{H}^{(gr)}\big>_{\!\!R}
&\approx
\frac{\hbar}{8\pi l_P^2\lambda}
\sum_{v}l_0^2\,N(v)
\frac{\sqrt{|p^{1}(v)\,p^{2}(v)|}}{l_0^2}
\,\tilde{c}_1(v)\,\tilde{c}_2(v)
\label{gravitational5}
\end{split}
\end{equation}
\begin{equation}
\begin{split}
{}_{\raisebox{-2.5pt}{\scriptsize$R$\!}}\big<\hat{H}^{(\phi)}_{kin}\big>_{\!\!R}
&\approx
\sum_{v}l_0^2\,N(v)
\frac{l_0^2}{\sqrt{|p^{1}(v)\,p^{2}(v)|}}\big(\hbar\pi^{(\phi)}_v\big)^{\!2}
\label{kinetic5}
\end{split}
\end{equation}
\begin{equation}
\begin{split}
{}_{\raisebox{-2.5pt}{\scriptsize$R$\!}}\big<\hat{H}^{(\phi)}_{der}\big>_{\!\!R}
&\approx
\sum_{v}l_0^2\,N(v)
\frac{\sqrt{|p^{1}(v)\,p^{2}(v)|}}{l_0^2}
\times\\
&\times
\Bigg\{
\frac{l_0^4\sqrt{|p^{1}(v)\,p^{2}(v)|}}{l_0^2\big(p^{(2)}(v)\big)^2}
\bigg(
\frac{e^{\rho(\hat{\phi}_{v\texttt{\,+}\vec{e}_1}\!-\hat{\phi}_{v})}\!-e^{\rho(\hat{\phi}_{v}-\hat{\phi}_{v\texttt{\,-}\vec{e}_1})}\!}{2\varepsilon\rho}
\bigg)^{\!\!2}
+\\
&+
\frac{l_0^4\sqrt{|p^{1}(v)\,p^{2}(v)|}}{l_0^2\big(p^{(1)}(v)\big)^2}
\bigg(
\frac{e^{\rho(\hat{\phi}_{v\texttt{\,+}\vec{e}_2}\!-\hat{\phi}_{v})}\!-e^{\rho(\hat{\phi}_{v}-\hat{\phi}_{v\texttt{\,-}\vec{e}_2})}\!}{2\varepsilon\rho}
\bigg)^{\!\!2}
\Bigg\},
\label{derivative5}
\end{split}
\end{equation}
\begin{equation}
\begin{split}
{}_{\raisebox{-2.5pt}{\scriptsize$R$\!}}\big<\hat{H}^{(\phi)}_{pot}\big>_{\!\!R}
&=
\sum_{v}l_0^2\,N(v)\sqrt{|p^{1}(v)\,p^{2}(v)|}\,{\scriptstyle V}\big(\phi(v)\big),
\label{potential5}
\end{split}
\end{equation}

Let us assume now to construct a proper semiclassical state, such that expectation values converge toward classical quantities. In the case of the gravitational term it has been already shown how the four-dimensional analog of the matrix element in \eqref{gravitational5} coincides with a semiclassical expression \cite{Alesci:2014uha}. This provides the correspondence principle \eqref{spin-momentum} for QRLG, which determines the value of the momentum corresponding to a classical configuration around which the state is peaked. Here, we assume the analogous relation for the matter field operators.

Then, introducing the inverse metric components --- see also Appendix~\eqref{AppendixB} --- $q^{11}=(p^1)^2/(l_0^4q^{\frac{3}{2}})$ and $q^{22}=(p^2)^2/(l_0^4q^{\frac{3}{2}})$, taking the continuum limit and replacing discrete variables with their continuous equivalents of classical fields, we obtain 
\begin{equation}
\begin{split}
{}_{\raisebox{-2.5pt}{\scriptsize$R$\!}}\big<\hat{H}\big>_{\!\!R}
&=
\!\int\!\!d^2x\,N(x)\,\sqrt{q(x)}
\bigg(
\frac{2}{\kappa\lambda}A^1_1(x)\,A^2_2(x)
+
\frac{1}{q(x)}\big(\pi^{(\phi)}(x)\big)^2\!
+
q^{11}(x)\big(\partial_1\phi(u)\big)^{2}\!+q^{22}(x)\big(\partial_2\phi(u)\big)^{2}\!
+
V\big(\phi(x)\big)
\!\bigg)\,.
\label{gravitational6}
\end{split}
\end{equation}
It is easy to see that the expression \eqref{gravitational6} reproduces the classical Hamiltonian \eqref{scalarconstraint} with the metric in the diagonal gauge.

\section{Conclusions and outlooks}\label{IV}
\noindent 
Moving from QRLG, we have shown how to implement Thiemann's procedure of Quantum Spin Dynamics in the three-dimensional case. Exemplifications encoded within the QRLG scheme, which resulted in the analytical expression for the eigenvalue of Hamiltonian constraint operator, were fundamental in deriving our results. The key feature we wish to emphasize here is that we addressed the quantum dynamics of a gravitational system coupled to matter. It is indeed the coupling to matter, as carrier of dynamics, that opened the pathway to the study of the quantum gravitational dynamics in three-dimensional models. One can easy extend our analysis so to encode vector fields, as well as different representations of the scalar field can be also deployed to the analysis of possible phenomenological implications.

Notice furthermore that considering a scalar field theory with generic potential $V(\phi)$ allows thorough a Legendre-Weil transform \cite{LW} to model any possible $f$(R) theory \cite{fR}, where the form of $f$ and $V$ are related. The scalar-tensor theory presented here can be thus seen as the quantization of a generic class of modified Einsteinian theories of gravity of $f$(R) type. Further possible applications point toward the deployment of $f$(R) gravitational models to analogue scenarios in condensed matter. The energy scale of the condensed matter system under scrutiny defines the effective Newton constant and the other related scales at the quantum level, while the curvature of the mimetic gravitational degrees of freedom encode the description of binding forces for lower spatial dimension lattice systems --- see \textit{e.g.} \cite{Kroener(1986),Kroner} and \cite{graphene}.

Over the paper, we derived the action of the Hamiltonian constraint operator for a system encoding a scalar field minimally coupled to gravity in three dimensions, using a new method that was developed in \cite{Bilski:2017ijd}. We also proposed how to define the potential operator for the scalar field, $\hat{V}\big(\phi(v)\big)$, and summarized its regularization  \cite{Bilski_Thesis,eigenvalues}. Probably our main result consists in having recovered that the Hamiltonian operator in \eqref{gravitational4} coincides with the one constructed in three-dimensional LQC \cite{Zhang:2014xqa,Ding:2016spw}. 

There are still open problems in the framework of QRLG. In particular we did not show yet how construct coherent states for the scalar field that will formally support the correspondence principle assumed in this article. Moreover, the different dependence of the metric determinant and the different coupling to the metric tensor for the cases in (\ref{kinetic5}), (\ref{derivative5}) and (\ref{potential5}) will generate different factors in the next to the leading order corrections for each of these terms. This suggests that the representation that has been chosen for this field is rather empty of phenomenological significance. It would be really difficult to extract measurable numbers from quantum corrections that scale in a different way for different terms of the same field. It opens a discussion how to restrict requirements imposed on the lattice representations of fields quantized via LQG techniques --- see \cite{Bilski_Thesis} for a sketch of such proposal.
	
\appendix
		%%%			%%%			%%%			%%%
\section{Thiemann's trick}\label{AppendixA}
		%%%			%%%			%%%			%%%
\noindent
In this appendix we repeat the procedure of lattice regularization \cite{Thiemann:2007zz} for the expressions that depend on the three-dimensional equivalents of the Ashtekar variables $(A_a^i,E_i^a)$.

First, using the Thiemann's trick \cite{Thiemann:1996aw}, we can derive the Poisson brackets between the two-dimensional equivalent of the Ashtekar connection $A_a^i$ \eqref{Ashtekarconnection} and the area $\mathbf{A}=\int\!d^2x\sqrt{q}$ of some surface $S$:
\begin{equation}\label{q_Etrick}
\mathbf{A}^{\!n}q_{ab}E^b_i
=
\frac{2}{n+2}\frac{\delta(\mathbf{A})^{n+2}}{\delta E^a_i}
=
\frac{2}{n+2}\frac{2}{\gamma\kappa}
\left\{A_a^i,\mathbf{A}^{\!n+2}\right\}.
\end{equation}
It is worth noting that this identity, together with \eqref{determinant} and \eqref{inversedeterminant}, provides the relations 
\begin{align}
\mathbf{A}^{\!n}\!E^a_i\,\mathbf{A}^{\!n}\!E^b_i
&=
\frac{1}{(n+1)^2}\frac{2^2}{(\gamma\kappa)^2}\tilde{\epsilon}^{ac}\,\tilde{\epsilon}^{bd}
\Big\{A_c^i,\mathbf{A}^{\!2(n+1)}\Big\}\Big\{A_d^i,\mathbf{A}^{\!2(n+1)}\Big\}\,,
\label{E^2trick}
\\
\mathbf{A}^{\!n}\,\mathbf{A}^{\!n}
&=
\frac{2^7}{(n+2)^4}\frac{2^4}{(\gamma\kappa)^4}\tilde{\epsilon}^{ac}\tilde{\epsilon}^{bd}
\Big\{A_a^i,\mathbf{A}^{\!n/2\,+1}\Big\}\Big\{A_b^i,\mathbf{A}^{\!n/2\,+1}\Big\}
\Big\{A_c^j,\mathbf{A}^{\!n/2\,+1}\Big\}\Big\{A_d^j,\mathbf{A}^{\!n/2\,+1}\Big\}.
\label{qtrick}
\end{align}

Next, expanding the SU$(2)$ holonomy along link $l^p$ outgoing from node $v$, and along loop $q\!\circlearrowleft\!r$ constructed from the anticlockwise sequence of links that starts start along link $l^q$ outgoing from the node $v$, we find the relations
\begin{subequations}
\begin{align}
h_p(v)
&=
1+\varepsilon A_p(v)+\mathcal{O}\big(\varepsilon^2\big)\,,
\\
h_{q\circlearrowleft r}(v)
&=
1+\frac{1}{2}\varepsilon^2F_{qr}(v)+\mathcal{O}\big(\varepsilon^4\big).
\end{align}
\end{subequations}
Here we used the following notation: $h_p(v):=h_{l^p}(v)$,
$h_{q\circlearrowleft r}(v)
=h_{l^q}(v)h_{l^r}(v)h_{l^q}^{-1}(v)h_{l^r}^{-1}(v)$
and $A_a:=A_a^i\tau_i$, $F_{ab}:=F^i_{ab}\tau_i$, with the $\mathfrak{su}(2)$ generators defined as $\tau_i=-\frac{i}{2}\sigma_i$, in which $\sigma_i$ denoting Pauli matrices.

Thus we obtain the formulas required for the implementation of the canonical quantization
\begin{subequations}
\begin{align}
\left\{A^i_a,\big(\mathbf{A}(S)\big)^n\right\}\delta^a_{l^p\!(v)}
&=
\frac{2}{\varepsilon}\,\text{tr}\Big(\tau^ih_p^{-1}(v)\Big\{h_p(v),\big(\mathbf{A}(S)\big)^n\Big\}\Big)
+\mathcal{O}(\varepsilon)\,,
\label{holonomyexpansion}
\\
F^3_{ab}\,\delta^a_{l^p\!(v)}\delta^b_{l^q\!(v)}
&=
-\frac{4}{\varepsilon^2}\,\text{tr}\Big(\tau^3h_{p\circlearrowleft q}(v)\Big)
+\mathcal{O}\big(\varepsilon^2\big),
\label{loopexpansion}
\end{align}
\end{subequations}
where, ``tr'' denotes the trace over $\mathfrak{su}(2)$ algebra indices and for simplicity we considered the fundamental representation. All the formulas above can be easly generalized to any irreducible representation of $\mathfrak{su}(2)$, possibly generating some additional factors in front of the right hand sides of \eqref{holonomyexpansion} and \eqref{loopexpansion}, which will be later canceled with analogous factors in \eqref{BCHtracetrick} and \eqref{BCHloop}.

\section{Flux operator in QRLG}\label{AppendixB}
\noindent
The canonical variables in four-dimensional QRLG are constructed from the diagonal connections and densitized dreibeins \cite{Alesci:2016gub}. The correspondence principle that links the spin $j^{(i)}$ at the quantum level with the classical canonical momentum $p^{(i)}$ reads
\begin{equation}\label{spin-momentum}
p^{(i)}=\kbar\Sigma^{(i)}_{v}\,.
\end{equation}
It is worth mentioning that this relation has been fixed in order to peak semiclassical states of QRLG \cite{Alesci:2014uha} around classical configurations, which is an assumption required by the complexifier method for constructing coherent states \cite{Thiemann:2000bw,Thiemann:2000ca}.

In order to reproduce all the standard four-dimensional identities for the reduced space coordinates $(c_{(i)},p^{(i)})$, we assume the correspondence principle \eqref{spin-momentum} and the canonical identity $\{c_i,p^j\}=\kappa\gamma\delta_i^j/2$,
with
$E_i^a=\tilde{p}^{(i)}\delta_i^a/l_0^2$,
and next we define
$q^{(4)}_{(i)(i)}=|p^1p^2p^3|/\big(l_0^2(p^i)^2\big)$,
obtaining
$q^{(4)}=|p^1p^2p^3|/l_0^6$.

Defining the three-dimensional equivalent of the determinant of the spatial metric, we find $q:=|p^1p^2|/l_0^4=q^{(4)}_{11}q^{(4)}_{22}$, having the solution $p^3=\sqrt{p^1p^2}$. Thus the metric tensor components read $q_{(i)(i)}=(|p^1p^2|)^{\frac{3}{2}}/\big(l_0^2(p^i)^2\big)$. However, the densitized triad written in terms of the reduced momenta reads $E_i^a\big|_{i=1,2}=(|p^1p^2|)^{\frac{3}{2}}\delta_i^a/(l_0^2p^{(i)})\big|_{i=1,2}$.

In order to recover the standard map between the reduced variables and the Ashtekar ones,
\begin{equation}
A^i_a=\frac{1}{l_0}\tilde{c}_{(i)}\delta^i_a\big|_{i=1,2}
\,,\quad
E_i^a\big|_{i=1,2}=\frac{1}{l_0^2}\tilde{p}^{(i)}\delta_i^a\big|_{i=1,2} \,,
\end{equation}
we define a new set of three-dimensional variables, namely
\begin{equation}
\tilde{c}_{(i)}=-\frac{4}{l_0}\frac{(p^{(i)})^2}{(|p^1p^2|)^{\frac{3}{4}}}c_{(i)}
\,,\quad
\tilde{p}^{(i)}=l_0\frac{(|p^1p^2|)^{\frac{3}{4}}}{p^{(i)}}
\,,\quad
\{\tilde{c}_i,\tilde{p}^j\}=\frac{1}{2}\kappa\gamma\delta_i^j\,.
\end{equation}
We can then express the reduced phase space variables in terms of the metric components, 
\begin{equation}\label{reduced-metric}
\tilde{c}_{(i)}\delta^i_a\big|_{i=1,2}=l_0\frac{\gamma}{N}\dot{\sqrt{|q_{(\!\:\!a\!\:\!)\!(\!\:\!a\!\:\!)}|}}\,\delta^i_a\big|_{i=1,2}
,\quad
|\tilde{p}^{(i)}|\delta_i^a\big|_{i=1,2}
=l_0^2\sqrt{\bigg|\frac{q_{11}\,q_{22}}{q_{(\!\:\!a\!\:\!)\!(\!\:\!a\!\:\!)}}\bigg|}\,\delta_i^a\big|_{i=1,2}.
\end{equation}

It is worth mentioning that the new set of variables $(\tilde{c}_{(i)},\tilde{p}^{(i)})$ is the solution of the constraint $q^{(4)}_{33}=1$, providing the following relation $q_{(i)(i)}=(\tilde{p}^1\tilde{p}^2)^2/\big(l_0^4(\tilde{p}^i)^2\big)$. This constraint can be interpreted as a condition that grades four-dimensional gravity in the ADM decomposition to its three-dimensional equivalent.

Knowing the relations between three-dimensional reduced variables and the Ashtekar variables, we can derive the action of the flux --- being a smeared dreibein, canonically conjugated to the holonomy of the connection $A^i_a$ assigned to the link of a length $\varepsilon$ --- on the reduced holonomy,
\begin{equation}\label{Schrodinger}
\hat{\tilde{p}}^i\big(S_v\big)h_{k}(v)
=-i\kbar\frac{\delta}{\delta c_{i}}e^{\varepsilon\tilde{c}_{(k)\!}(v)\tau^{k}\!/l_0}
=-i\kbar\frac{\varepsilon}{l_0}\tau^ih_{k}(v),
\end{equation}
where
$\hat{\tilde{p}}^i\big(S\big):=\frac{1}{2}\Big(\hat{E}_{(i)}\big({S}^{p}\big)+\hat{E}_{(i)}\big({S}^{-p}\big)\Big)\delta^i_p$ 
and ${S}^p$ is the surface normal to link $l^p$, with fiducial area $l_0^2$.
We also used the exact formula for the U$(1)$ holonomy,
$h_{p}:=e^{\pm\varepsilon A_{(i)\!}\tau^i}\delta^i_{p\,}$, where the sign of the exponent depends on the orientation of the link $l^p$ with respect to the chosen frame --- here we are assuming the Cartesian right-oriented reference frame, with $x$ and $y$ directions parallel to the lattice links.

We can easily transpose these findings into the action on the reduced Wigner matrix carrying generator $\tau^{(l)}$ of the irreducible representation $j_l$, gauge fixed to diagonal along the link $l$,
\begin{equation}
{}^{l\!}D^{j_l}_{m_lm_l}(h_l)=\left<m_l,\vec{u}_l\middle|D^{j_l}(h_l)\middle|m_l,\vec{u}_l\right>
,\quad
m_l=\pm j_l.
\end{equation}
We then obtain 
\begin{equation}
\hat{\tilde{p}}^i\big(S\big)\,{}^{l\!}D^{j_l}_{m_lm_l}(h_l)
=\frac{\kbar}{2}\big(j^{(i)}_{v}+j^{(i)}_{v-\vec{e}_i}\big)D^{j_l}_{m_lm_l}(h_l)
=\kbar\Sigma^{(i)}_{v}
D^{j_l}_{m_lm_l}(h_l)
,\quad
l=l^i.
\end{equation}

		%%%			%%%			%%%			%%%
\section{QRLG trace action}\label{AppendixC}
		%%%			%%%			%%%			%%%
\noindent
In this appendix we present new methods to computate the action of the traces of the operators
$\text{tr}\big(\tau^i\hat{h}_p^{-1}\hat{\mathbf{A}}^n(v)\,\hat{h}_p\big)$
and
$\text{tr}\big((\hat{h}_{i\circlearrowleft j}-\hat{h}_{i\circlearrowleft j}^{-1})\tau^3\big)$.
In this derivation we follow the method proposed in \cite{Bilski:2017ijd}, which is based on the BCH expansion of $\mathfrak{su}(2)$-valued eigenvalues of the considered traces of operators. An alternative procedure is presented in \cite{Alesci:2013xd,Alesci:2014uha}.

The action of the first trace of operators reads
\begin{equation}
\begin{split}
\label{BCHtracetrick}
\text{tr}
\Big(
\tau^i\hat{h}_j^{-1}\,\hat{\mathbf{A}}^n(v)\,\hat{h}_j
\!\Big)
\!\ket{\Gamma;j_l,i_v}_{\!R}
=&\ 
\,\text{tr}
\Bigg(\!
\tau^i\,h_j^{-1}\big(\mathbf{A}_{v}\big)^{\!n}\bigg(\mathds{1}+\frac{i\varepsilon}{2l_0\Sigma_v^{(j)}}\tau^j\Bigg)^{\!\!\!\frac{n}{2}}h_j
\!\Bigg)
\!\ket{\Gamma;j_l,i_v}_{\!R}
\\
=&\,
-\frac{i\varepsilon
n\,\delta^i_j}{2^3l_0\Sigma^{(j)}_v}\big(\mathbf{A}_{v}\big)^{\!n}
\sum_{r=1}\frac{2^{5-4r}}{n}\bigg(\frac{\varepsilon}{l_0\Sigma_v^{(j)}}\!\bigg)^{\!\!2(r-1)}\binom{n/2}{2r-1}
\!\ket{\Gamma;j_l,i_v}_{\!R}
\\
=&\,
-\frac{i\varepsilon
n\,\delta^i_j}{2^3l_0\Sigma^{(j)}_v}\big(\mathbf{A}_{v}\big)^{\!n}
\Big(1+\mathcal{O}(\varepsilon^2)\Big)
\!\ket{\Gamma;j_l,i_v}_{\!R},
\end{split}
\end{equation}
where for simplicity we assumed the $\mathfrak{su}(2)$ generators to be in the fundamental representation. It is worth mentioning that this result can generalized for any irreducible representation of $\mathfrak{su}(2)$, possibly modifying the factor in front of the second and the third line of \eqref{BCHtracetrick}, which will be finally simplified with an analogous factor coming from \eqref{holonomyexpansion}.

Concerning the second trace of operators, using the BCH formula and gathering all the terms coming from the expansion of the exponentiated operators in the definition of the U$(1)$ holonomy, we find
\begin{equation}\label{BCHloop}
\begin{split}
\,\text{tr}\Big(\big(\hat{h}_{i\circlearrowleft j}-\hat{h}_{i\circlearrowleft j}^{-1}\big)\tau^3\Big)
\ket{c_v;j_l,i_v}_{\!R}
=&\,-\epsilon_{ij3}\sin\Big(\frac{\varepsilon}{l_0}c_{(i)\!}(v)\Big)\sin\Big(\frac{\varepsilon}{l_0}c_{(j)\!}(v)\Big)
\!\ket{\Gamma;j_l,i_v}_{\!R}
\\
=&\,-\epsilon_{ij3}c_{(i)\!}(v)\,c_{(j)\!}(v)\frac{\varepsilon^2}{l_0^2}\Big(1+\mathcal{O}(\varepsilon^2)\Big)
\!\ket{\Gamma;j_l,i_v}_{\!R}.
\end{split}
\end{equation}
Notice that the non-expanded expression above reproduces the same gravitational contribution to the Hamiltonian constraint operator, as the one provided by the regulated operator in three-dimensional LQC \cite{Zhang:2014xqa,Ding:2016spw}. Besides that, similarly as in the case of the previous formula, if we consider an irreducible representation of $\mathfrak{su}(2)$ other than the fundamental one, the possible change in the overall factor, coming from the contraction of generators, will cancel the factor that will appear from the formula \eqref{holonomyexpansion}.

\vspace{1cm}

{\it{\textbf{Acknowledgments}}}\\
\noindent
The authors wish to thank S.~Brahma for useful discussions.
A.M. wishes to acknowledge support by the Shanghai Municipality, through the grant No. KBH1512299, and by Fudan University, through the grant No. JJH1512105. 	
	
%\newpage

	%%%%%%%%%	%%%%%%%%%	%%%%%%%%%	%%%%%%%%%
\end{document}